\documentclass[12pt]{article}
\makeatletter
\@addtoreset{equation}{section}

\makeatother

\begin{document}

\begin{titlepage}
\null
\begin{flushright}
UT-Komaba/03-17
\\
hep-th/0309265
\\
September, 2003
\end{flushright}

\vskip 1.5cm
\begin{center}

  {\Large \bf Towards Noncommutative 
Integrable Equations\footnote{Talk given 
by K.T. at the fifth international conference
on Symmetry in Nonlinear Mathematical Physics, 
Kiev, Ukraine, 23-29 June 2003.}
}

\lineskip .75em
\vskip 1.5cm
\normalsize

 {\large Masashi Hamanaka\footnote{e-mail: 
hamanaka@hep1.c.u-tokyo.ac.jp}}
and
 {\large Kouichi Toda\footnote{e-mail:
kouichi@yukawa.kyoto-u.ac.jp}}

\vskip 1.5cm

  ${}^1${\it Institute of Physics, University of Tokyo, Komaba,\\
              Meguro-ku, Tokyo 153-8902, Japan}

\vskip 0.5cm

  ${}^2${\it Department of Mathematical Physics,
Toyama Prefectural University,\\
Toyama 939-0398, Japan}

\vskip 1.5cm

{\bf Abstract}

\end{center}

We study the extension of integrable equations
which possess the Lax representations
to noncommutative spaces.
We construct various noncommutative Lax equations
by the Lax-pair generating technique and the Sato theory.
The Sato theory has revealed essential aspects of the integrability
of commutative soliton equations 
and the noncommutative extension is worth studying.
We succeed in deriving various noncommutative
hierarchy equations in the framework of the Sato theory,
which is brand-new.
The existence of the hierarchy would
suggest a hidden infinite-dimensional symmetry
in the noncommutative Lax equations. 
We finally show that
a noncommutative version of Burgers equation
is completely integrable because it is
linearizable via noncommutative Cole-Hopf transformation.
These results are expected to lead to the completion of 
the noncommutative Sato theory.

\end{titlepage}
\clearpage
\baselineskip 6.5mm

\section{Introduction}

The extension of ordinary integrable systems
to noncommutative (NC) spaces is
one of hot topics in the recent study of
integrable systems \cite{Toda:CFZ}-\cite{Toda:WaWa}.
NC extension in gauge theories corresponds to
the presence of background magnetic fields
and leads to the discovery of many new physical objects
and successful applications to
string theories \cite{Toda:NC}.
In particular, NC (anti-)self-dual Yang-Mills (YM)
equations are integrable and important \cite{Toda:YM}.

On the other hand,
many typical integrable equations
such as 
the Kadomtsev-Petviashvili (KP) equation \cite{Toda:KaPe},
the Korteweg-de Vries (KdV) equation \cite{Toda:KdV},
and the Boussinesq equation \cite{Toda:Boussinesq}, 
contain no gauge field and
the NC extension of them perhaps might have  
no physical picture.
NC extension of $(1+1)$-dimensional nonlinear equations
introduces infinite number of time derivatives
and it becomes very hard to define the integrability.
Nevertheless, some of them 
actually possess integrable properties, 
such as the existence of infinite number 
of conserved quantities \cite{Toda:DiMH2, Toda:GrPe}.
Furthermore, a few of them can be derived from 
NC (anti-)self-dual YM equations
by suitable reductions \cite{Toda:Legare, Toda:HaTo2}.
This fact may give some physical meanings and good properties
to the lower-dimensional NC field equations.
Now it is time to study in detail
whether they are actually integrable or not.

In this article,
we present various NC equations
which possess the Lax representations \cite{Toda:Lax}.
We mainly discuss the Lax-pair generating technique
and applications of the Sato theory.
The Sato theory is one of the most beautiful
soliton theories and reveals various 
integrable aspects of soliton equations, such as
the existence of multi-soliton solutions,
the structure of the solution space
and the hidden symmetry of them.
Hence NC extension of the Sato theory is worth studying
and the present discussions on it are all new.
Here we prove the existence of various NC hierarchy equations, 
which would suggest hidden infinite-dimensional
symmetries of NC equations.
Finally we discuss the integrability
of NC $(1+1)$-dimensional Burgers equation and 
prove that it is completely integrable 
due to the linearizability.

\section{Noncommutative Field Equations}

NC spaces are defined
by the noncommutativity of the coordinates:
\begin{eqnarray}
\label{Toda:nc_coord}
[x^i,x^j]=i\theta^{ij},
\end{eqnarray}
where $\theta^{ij}$ are real constants and  
called the {\it NC parameters}.

NC field theories can be defined by
the replacement of ordinary products of fields 
in the commutative theories with 
the {\it star-product}.
The star-product is defined for ordinary fields 
on commutative spaces and explicitly given by 
\begin{eqnarray}
f\star g(x)&:=&
\exp{\left(\frac{i}{2}\theta^{ij}\partial^{(x^\prime)}_i
\partial^{(x^{\prime\prime})}_j\right)}
f(x^\prime)g(x^{\prime\prime})\Big{\vert}_{x^{\prime}
=x^{\prime\prime}=x,}
\label{Toda:star}
\end{eqnarray}
where $\partial_i^{(x^\prime)}:=\partial/\partial x^{\prime i}$
and so on.
The star-product has associativity: $f\star(g\star h)=(f\star g)\star h$
and returns back to the ordinary product 
in the commutative limit: $\theta^{ij}\rightarrow 0$.
The modification of the product  makes the ordinary 
spatial coordinate ``noncommutative,'' 
that is, $[x^i,x^j]_\star:=x^i\star x^j-x^j\star x^i=i\theta^{ij}$.

We note that the fields themselves take c-number values 
as usual and the differentiation and the integration for them 
are well-defined as usual.
A nontrivial point is that
NC field equations contain infinite number of
derivatives in the nonlinear terms.
Hence the integrability of the equations
are not so trivial as in commutative cases.
For detailed discussion on it, see \cite{Toda:HaTo2}.

\section{Noncommutative Lax Equations}

A given NC differential equation is said to have
the Lax representation
if there exists a suitable pair of operators $(L,B)$
so that the following equation
\begin{eqnarray}
 \left[\partial_{t}-B, L\right]_\star=0,
\label{Toda:lax}
\end{eqnarray}
is equivalent to the given NC differential equation.
Here the star-product does not affect the derivative operator,
for example, $\partial_t \star \partial_x = \partial_t \partial_x$.
The pair of operators $(L,B)$ and the equation (\ref{Toda:lax})
are called the {\it Lax pair} and the {\it NC Lax equation},
respectively. 

On NC spaces, the meaning of Lax representations
would be vague \cite{Toda:HaTo2}.
However, they actually have close connections with
the bi-complex method \cite{Toda:DiMH}
which leads to infinite number of conserved quantities,
and the (anti)-self-dual YM equation
which is integrable in the context of
twistor descriptions and ADHM constructions \cite{Toda:YM}.


Now let us construct NC Lax equations
by the {\it Lax-pair generating technique}.
The technique is a method to
find a corresponding $B$-operator for a given $L$-operator
and based on the following ansatz for the $B$-operator
\begin{eqnarray}
\label{Toda:ansatz}
B=\partial_i^n L^m +B^\prime.
\end{eqnarray}
Then the problem reduces to that for the $B^\prime$-operator
which is determined by hand so that the
Lax equation should be a differential equation
without bare differential $\partial_i$.

In order to explain the steps, for example, let us
consider the KdV equation on NC $(1+1)$-dimensional
space-time where the coordinate and the noncommutativity
are denoted by $(x,t)$ and $[t,x]=i\theta$, respectively.

\vspace{1mm}
\noindent
\underline{NC KdV equation \cite{Toda:DiMH2,Toda:Legare,Toda:Toda,Toda:HaTo}}
\noindent
\vspace{2mm}

The $L$-operator for KdV equation is given
by $L_{\scriptsize\mbox{KdV}}=\partial_x^2+u(x,t)$.
The ansatz for the operator $B$ is of the following type:
\begin{eqnarray}
B_{\scriptsize\mbox{KdV}}=\partial_x L_{\scriptsize\mbox{KdV}} +B^\prime
=\partial^3_x+u\partial_x+u_x+B^\prime,
\label{ansatz}
\end{eqnarray}
where $u_x:=\partial u/\partial x$.
The Lax equation (\ref{Toda:lax}) leads to the equation for
the unknown operator $B^\prime$:
\begin{eqnarray}
[\partial_x^2+u,B^\prime]=u_x\partial_x^2+u_x\star u-u_t.
\end{eqnarray}
Here the bare derivative term 
$u_x\partial_x^2$ is troublesome.
In order to delete it,
let us take the following ansatz for $B^\prime$:
\begin{eqnarray}
B^\prime=X \partial_x+Y,
\end{eqnarray}
where $X$ and $Y$ are polynomial of $u, u_x, u_t, 
u_{xx}:=\partial^2 u/\partial x^2$ etc.
The Lax equation (\ref{Toda:lax}) reduces to
\begin{eqnarray}
 (2X-u)_x\partial_x^2+(X_{xx}+[u,X]+2Y_x)\partial_x
 +(Y_{xx}+[u,Y]-X\star u_x+u_t-u_x\star u)=0.
\end{eqnarray}
The condition that
the coefficients of $\partial_x^2$ and $\partial_x$ 
should vanish yields differential equations for $X$ and $Y$, 
which are easily solved by
\begin{eqnarray}
X=\frac{1}{2}u,~~~Y=-\frac{1}{4}u_x.
\end{eqnarray}
Now the Lax equation (\ref{Toda:lax}) 
becomes a differential equation,
that is, the NC KdV equation:
\begin{eqnarray}
\label{Toda:ncKdV}
u_t=\frac{1}{4}u_{xxx}+\frac{3}{4}(u\star u)_x,
\end{eqnarray}
where $(u\star u)_x:= u_x\star u+u\star u_x$.
The nonlinear term becomes symmetric
and the equation shows just a conservation law.
In the star-product formalism,
the spatial integration is well-defined.
Hence the spatial integrations of current densities
are conserved quantities as in commutative cases \cite{Toda:HaTo2}.
Here $Q:=\int dx u$ is conserved, that is, $\partial_t Q=0$.

\vspace{2mm}


In this way, we can generate wide class of Lax equations
on NC $(2+1)$ and $(1+1)$-dimensional space-times. 
In particular, this method is suitable for
higher-dimensional extension both on commutative spaces
\cite{Toda:ToYu} and NC spaces \cite{Toda:Toda}.
Here we present 
some results of $(2+1)$-dimensional NC Lax equations
where the coordinate are denoted by $(x,y,t)$.
The noncommutativity is basically introduced
in space-space directions: $[x,y]=i\theta$. 
For more discussions and examples, 
see \cite{Toda:Toda, Toda:HaTo}.

\vspace{1mm}
\noindent
\underline{NC Kadomtsev-Petviashvili (KP) 
equation \cite{Toda:Paniak, Toda:Toda, Toda:HaTo}}
\noindent
\begin{eqnarray}
\label{Toda:ncKP}
u_t=\frac{1}{4}u_{xxx}+\frac{3}{4}(u\star u)_x
+\frac{3}{4}\partial_x^{-1}u_{yy}
+\frac{3}{4}[u,\partial_x^{-1} u_y]_\star,
\end{eqnarray}
where $\partial_x^{-1}f(x):=\int^x dx^\prime f(x^\prime)$.
The Lax pair is given by
\begin{eqnarray}
L_{\scriptsize\mbox{KP}}&=&\partial_x^2+u(x,y,t)+\partial_y
=:L_{\scriptsize\mbox{KP}}^\prime+\partial_y,\nonumber\\
B_{\scriptsize\mbox{KP}}&=&\partial_x L_{\scriptsize\mbox{KP}}^\prime
+X\partial_x+Y=\partial_x^3+\frac{3}{2}u\partial_x+\frac{3}{4}u_x
-\frac{3}{4}\partial_x^{-1}u_y.
\end{eqnarray}
There is seen to be a nontrivial 
deformed term $[u,\partial_x^{-1} u_y]_\star$
in the equation (\ref{Toda:ncKP}), which vanishes 
in the commutative limit.
This reduces to the NC KdV equation (\ref{Toda:ncKdV}) 
setting ``$\partial_y=0$''
and taking space-time noncommutativity.

\vspace{1mm}
\noindent
\underline{NC Bogoyavlenski-Calogero-Schiff (BCS) 
equation \cite{Toda:Toda, Toda:HaTo}}
\noindent
\begin{eqnarray}
u_t=\frac{1}{4}u_{xxy}+\frac{1}{2}(u\star u)_y
+\frac{1}{4}u_x\star (\partial_x^{-1} u_y)
+\frac{1}{4}(\partial_x^{-1}u_{y})\star u_x
+\frac{1}{4}[u,\partial_x^{-1}[u,\partial_x^{-1}u_{y}]_\star]_\star,
\end{eqnarray}
whose Lax pair and the ansatz are
\begin{eqnarray}
L_{\scriptsize\mbox{BCS}}&=&\partial_x^2+u(x,y,t),\nonumber\\
B_{\scriptsize\mbox{BCS}}&=&\partial_y L_{\scriptsize\mbox{BCS}}
+X\partial_x+Y\nonumber\\
&=&\partial_x^2\partial_y+u\partial_y
+\frac{1}{2}(\partial_x^{-1}u_y)\partial_x
+\frac{3}{4}u_y
-\frac{1}{4}\partial_x^{-1}[u,\partial_x^{-1}u_y]_\star.
\end{eqnarray}
This time, a nontrivial term is found even in the $B$-operator.
In commutative limit, this coincides with 
the BCS equation \cite{Toda:BCS} which has
multi-soliton solutions \cite{Toda:YTSF}.
This reduces to the NC KdV equation (\ref{Toda:ncKdV}) 
setting $x=y$ and taking space-time noncommutativity.

\section{Noncommutative Hierarchy Equations and the Sato Theory}

In this section, we derive various
noncommutative hierarchy equations
in the framework of the Sato theory \cite{Toda:SaSa}
introducing the pseudo-differential operator.

Let us introduce the following Lax operator
as a (first-order) pseudo-differential operator:
\begin{eqnarray}
 L = \partial_x + u_2 \partial_x^{-1} 
 + u_3 \partial_x^{-2} + u_4 \partial_x^{-3} + \cdots,~~~
 u_k=u_k(t_1,t_2,t_3,~\ldots).
\end{eqnarray}
The action of the operator $\partial_x^n$ on
a multiplicity operator $f$ is given by
\begin{eqnarray}
 \partial_x^{n}\cdot f:=\sum_{i\geq 0}
\left(\begin{array}{c}n\\i\end{array}\right)
(\partial_x^i f)\partial^{n-i}:=
\sum_{i\geq 0}
\frac{n(n-1)\cdots (n-i+1)}{i(i-1)\cdots 1}
(\partial_x^i f)\partial^{n-i}.
\end{eqnarray}
We note that the definition 
can be extended to negative $n$.
The examples are,
\begin{eqnarray*}
 \partial_x^{-1}\cdot f&=& 
f\partial_x^{-1}-f_x\partial_x^{-2}+f_{xx}\partial_x^{-3}-\cdots,\\
 \partial_x^{-2}\cdot f&=& 
f\partial_x^{-2}-2f_x\partial_x^{-3}+3f_{xx}\partial_x^{-4}-\cdots,\\
 \partial_x^{-3}\cdot f&=& 
f\partial_x^{-3}-3f_x\partial_x^{-4}+6f_{xx}\partial_x^{-5}-\cdots,
\end{eqnarray*}
where $\partial_x^{-1}$ in the RHS
acts as an integration operator $\int^x dx$.
Products of pseudo-differential operators
are also well-defined and the total set 
of pseudo-differential operators forms 
an operator algebra.
For more on 
pseudo-differential operators 
and the Sato theory, see e.g. \cite{Toda:DJM}.

The Lax representation for a hierarchy
in Sato's framework is defined as
\begin{eqnarray}
 \left[\partial_{t_m}-B_m, L\right]_\star=0,~~~m=1,2,\cdots,
\label{Toda:lax_sato}
\end{eqnarray}
where $B_m$ is given here by
\begin{eqnarray}
 B_m
:=(\underbrace{L\star \cdots \star L}_{m{\scriptsize\mbox{ times}}})_{\geq 0}
=:(L^m)_{\star\geq 0}.
\end{eqnarray}
The suffix ``$\geq 0$'' represents the positive and 0-th 
power part of $L^m$. The examples are
\begin{eqnarray}
B_1=\partial_x,~~~B_2=\partial_x^2 + 2u_2,~~~
B_3=\partial_x^3 + 3u_2\partial_x + 3(u_3+u_{2x}),\cdots.
\end{eqnarray}

The noncommutativity is introduced for
infinite number of ``time variables''
$(t_1,t_2,\ldots)$.
As it can be taken arbitrarily,
we do not fix the noncommutativity here.

\vspace{1mm}
\noindent
\underline{NC KP Hierarchy}
\noindent
\vspace{2mm}

The hierarchy (\ref{Toda:lax_sato}) 
gives rise to NC KP hierarchy
which contains the NC KP equation (\ref{Toda:ncKP}).
The coefficients of each powers of (pseudo-)differential
operators yield infinite series of NC
``evolution equations,'' 
that is,
for $m=1$
\begin{eqnarray}
\partial_x^{1-k})~~~ u_{kt_1}=u_{kx},~~~k=2,3,\cdots 
~~~\Rightarrow~~~t_1\equiv x,
\end{eqnarray}
for $m=2$
\begin{eqnarray}
\label{Toda:KP_hie}
\partial_x^{-1})~~~u_{2t_2}&=&u_{2xx}+2u_{3x},\nonumber \\
\partial_x^{-2})~~~
u_{3t_2}&=&u_{3xx}+2u_{4x}+2u_2\star u_{2x} +2[u_2,u_3]_\star,\nonumber \\
\partial_x^{-3})~~~
u_{4t_2}&=&u_{4xx}+2u_{5x}+4u_3\star u_{2x}-2u_2\star u_{2xx}
+2[u_2,u_4]_\star,\cdots
\end{eqnarray}
and  for $m=3$
\begin{eqnarray}
\partial_x^{-1})~~~
u_{2t_3}&=&u_{2xxx}+3u_{3xx}+3u_{4x}+3u_{2x}\star u_2+3u_2\star u_{2x},
\\
\partial_x^{-2})~~~
u_{3t_3}&=&u_{3xxx}+3u_{4xx}+3u_{5x}+6u_{2}\star u_{3x}+3u_{2x}\star u_3 
+3u_3\star u_{2x}+3[u_2,u_4],\cdots.\nonumber
\end{eqnarray}
These just imply the NC KP equation (\ref{Toda:ncKP}) 
with $2u_2\equiv u, t_2\equiv y,t_3\equiv t$.
Important point is that infinite kind of fields $u_3, u_4, u_5,\ldots$
are represented in terms of one kind of field  $2u_2\equiv u$
as is seen in Eq. (\ref{Toda:KP_hie}).
This guarantees the existence of NC KP hierarchy
and the infinite differential equations
are called the {\it NC (KP) hierarchy equations}.

\vspace{2mm}

Putting the constraint $L^l=B_l$
on the NC KP hierarchy (\ref{Toda:lax_sato}),
we get infinite set of NC hierarchies.
We can easily show
\begin{eqnarray}
\frac{\partial u}{\partial t_{Nl}}=0,~~~
N=1,2,\ldots
\end{eqnarray}
because of $B_{Nl}=L^{Nl}$.
The reduced NC hierarchy is called the {\it l-reduction}
of NC KP hierarchy.

\vspace{1mm}
\noindent
\underline{NC KdV Hierarchy (2-reduction of NC KP)}
\noindent
\vspace{2mm}

Taking the constraint $L^2=B_2$,
we get the KdV hierarchy. The Lax equation
\begin{eqnarray}
 \frac{\partial u}{\partial t_m}=\left[B_m, L^2\right]_\star,
\end{eqnarray}
gives rise to the $m$-th NC KdV hierarchy equation
which becomes, for example, the (third) NC KdV 
equation (\ref{Toda:ncKdV}) with $t_3\equiv t$
and the 5-th NC KdV equation \cite{Toda:Toda}
\begin{eqnarray}
 u_{t_5}&=&\frac{1}{16}u_{xxxxx}+\frac{5}{16}(u\star u_{xxx}+u_{xxx}\star u)
+\frac{5}{8}(u_x\star u_x+u\star u\star u)_x.
\end{eqnarray}
In the Lax-pair generating technique,
the $B_m$-operator is given by the ansatz
$B_m=\partial_x^{m-2}L_{\scriptsize\mbox{KdV}}+B^\prime$.

\vspace{1mm}
\noindent
\underline{NC Boussinesq Hierarchy (3-reduction of NC KP)}
\noindent
\vspace{2mm}

The 3-reduction  $L^3=B_3$ yields the NC Boussinesq hierarchy
which contains the NC Boussinesq equation \cite{Toda:Toda}:
\begin{eqnarray}
 u_{tt}=\frac{1}{3}u_{xxxx}+(u \star u)_{xx}
+([u,\partial_x^{-1}u_t]_{\star})_x.
\end{eqnarray}

In this way, we can generate 
infinite set of the $l$-reduced NC hierarchies.
Furthermore, if we take other set-up for
the definition of pseudo-differential operator $L$
and ``time-evolution'' operator $B_m$,
we can get many other hierarchies as follows.

\vspace{1mm}
\noindent
\underline{NC modified KdV (mKdV) Hierarchy}
\noindent
\vspace{2mm}

If we take
\begin{eqnarray}
 L = \partial_x + u_1
+ u_2 \partial_x^{-1} 
 + u_3 \partial_x^{-2} + \cdots,~~~
 B_m=(L^m)_{\star\geq 1},
\end{eqnarray}
and put the constraint $L^2=B_2$,
infinite kind of fields $u_2,u_3,\ldots$ are 
represented in terms of one kind of field $2u_1\equiv v$.
We can easily see $\partial_{t_{2N}} v=0$.
The set of Lax equations
\begin{eqnarray}
[B_m-\partial_{t_m}, L^2]_\star=0
\end{eqnarray}
yields NC mKdV hierarchy
which implies the NC mKdV equation for $m=3$
with $t_3\equiv t$:
\begin{eqnarray}
\label{Toda:ncmKdV}
\frac{\partial v}{\partial t}
=\frac{1}{4}v_{xxx}-\frac{3}{8}v\star v_x\star v+\frac{3}{8}
[v,v_{xx}]_\star.
\end{eqnarray}
We note that the NC mKdV equation (\ref{Toda:ncmKdV}) 
is different from that in \cite{Toda:DiMH2} 
from NC KdV equation (\ref{Toda:ncKdV}) 
via NC Miura map $u=-v_x-v\star v$.
The NC mKdV hierarchy can be considered as the 2-reduction
of NC mKP hierarchy. However the existence of the NC mKP hierarchy 
seems to be nontrivial because it can not be represented
in terms of one kind of field. 
(The NC mKP equation given in early versions of \cite{Toda:HaTo} 
 is incorrect.)
For the same reason, the higher reduction
of the NC mKP hierarchy seems to be hard to obtain.

\vspace{1mm}
\noindent
\underline{NC Burgers Hierarchy \cite{Toda:HaTo2}}
\noindent
\vspace{2mm}

If we take
\begin{eqnarray}
 L = \partial_x + u_1 + u_2 \partial_x^{-1} 
 + u_3 \partial_x^{-2} + \cdots,~~~
 B_m=(L^m)_{\star\geq 1},
\end{eqnarray}
and put the constraint $L=\partial_x+v$,
the hierarchy 
\begin{eqnarray}
 \partial_{t_m}v=[B_m, L]_\star
\end{eqnarray}
yields the NC Burgers hierarchy
which implies the NC Burgers equation
\begin{eqnarray}
\label{Toda:burgers}
 \frac{\partial v}{\partial t_2} 
= [B_2,L]_\star
= [\partial_x^2+2v\partial_x, \partial_x +v]_\star
 =v_{xx}+2v\star v_x.
\end{eqnarray}
and the third order NC Burgers equation
\begin{eqnarray}
 \frac{\partial v}{\partial t_3}
=[B_3,L]_\star
 =v_{xxx}+3v\star v_{xx}+3v_x\star v_x +3v\star v\star v_{x}
\end{eqnarray}
and so on. The nonlinear terms are not symmetric,
which will be proved to be a key point in the linearization
in the next section.
The $B_m$-operator is given in the Lax-pair generating technique
by $B_m=\partial_x^{m-1} L+B^\prime$. This time, 
the generated equations contain some parameters 
and thus covers wider class of Lax equations \cite{Toda:HaTo2}.

\section{Integrability of Noncommutative Burgers Equation}

In this section, we discuss the integrability
of the Burgers equation (\ref{Toda:burgers})
on NC $(1+1)$-dimensional space-time
where the coordinate and the noncommutativity
are denoted by $(x,t)\equiv (t_1,t_2)$ 
and $[t,x]=i\theta$, respectively.

In commutative case, it is well known that
the Burgers equation \cite{Toda:Burgers}
is linearized by the
Cole-Hopf transformation \cite{Toda:CoHo}.
The discussion can be extended to
noncommutative case \cite{Toda:HaTo2,Toda:MaPa}.
NC Burgers equation (\ref{Toda:burgers})
can be linearized by
the following noncommutative analogue
of the Cole-Hopf transformation
\begin{eqnarray}
 v=\psi^{-1}\star\psi_x.
\end{eqnarray}
The linearized equation is a (NC)
diffusion equation
\begin{eqnarray}
\label{Toda:diffusion}
 \psi_t=\psi_{xx}.
\end{eqnarray}
The naive solution of 
the NC diffusion equation (\ref{Toda:diffusion}) is
\begin{eqnarray}
\label{Toda:naive_sol}
\psi(t,x)=1+\sum_{i=1}^N h_i e^{k_i^2t}\star e^{\pm k_i x}
=1+\sum_{i=1}^N h_i e^{\frac{i}{2}k_i^3\theta}e^{k_i^2t\pm k_i x},
\end{eqnarray}
where $h_i, k_i$ are complex constants.
In the commutative limit,
this reduces to the $N$-shock wave solution 
in fluid dynamics.
The final form in (\ref{Toda:naive_sol})
shows that the $N$-shock wave solution
is deformed by $e^{\frac{i}{2}k_i^3\theta}$ 
due to the noncommutativity.
The explicit representation in terms of $v$
is hard to obtain because the derivation of $\psi^{-1}$
is nontrivial.
However we can discuss the asymptotic behaviors
at $t\rightarrow \pm \infty$ and actually see 
the effect of the NC deformation.
In fact, the exact solutions for $N=1,2$ 
are obtained in \cite{Toda:MaPa} 
and nontrivial effects of the NC-deformation
are reported.
We note that NC one shock-wave solutions
can always reduce to the commutative ones
because $f(t-x)\star g(t-x)=f(t-x)g(t-x)$ \cite{Toda:HaTo2}.

The results show that the NC Burgers equation 
(\ref{Toda:burgers})
is completely integrable even though the
NC Burgers equation contains infinite number
of time-derivatives in the nonlinear term.
The linearized equation is a differential equation
of first order with respect to time
and the initial value problem is well-defined.
This is a surprising result.
The (NC) diffusion equation
can be solved for arbitrarily boundary conditions 
by the Fourier transformation.
Furthermore, we note that 
the form of the nonlinear term in the NC Burgers
equation (\ref{Toda:burgers}) is crucial for the linearization.
If it becomes symmetric like the NC KdV equation (\ref{Toda:ncKdV}),
the linearization is proved to 
be impossible \cite{Toda:HaTo2}.

\section{Conclusion and Discussion}

In this article,
we presented various NC Lax equations
and proved the existence of many NC hierarchies.
The NC extension of the Sato theory is new.
We also confirmed that
NC Burgers equation is linearizable and 
completely integrable
even though it is a differential equation of infinite order
with respect to time.
The linearized equation is a (NC) diffusion
equation and can be solved in usual ways.

NC extension of the Ward conjecture \cite{Toda:Ward}
would be very interesting \cite{Toda:HaTo}
though we have omitted it in this article
because of limitations of space.
Some NC equations are actually derived from
NC (anti-)self-dual YM equations by reduction
\cite{Toda:Legare,Toda:HaTo2}
and embedded \cite{Toda:LPS,Toda:LePo, Toda:LPS2}
in $N=2$ string theory \cite{Toda:OoVa}.
This guarantees that NC integrable equations
would have a physical meaning and 
might be helpful to understand new aspects of 
the corresponding string theory.

The next step is NC extension of 
Hirota's bilinearization \cite{Toda:Hirota}.
This could be realized as a simple generalization of
the Cole-Hope transformation whose extension
to NC spaces are already successful as shown in Sec. 5.
Hirota's bilinearization leads to
the theory of tau-functions.
The Sato theory is based on
the existence of hierarchies and tau-functions.
We have just revealed the existence of hierarchies
and the completion of the NC Sato theory
would be drawing near at hand.

\subsection*{Acknowledgements}

We would like to thank M.~Kato, I.~Kishimoto, A.~Nakamula
and T.~Tsuchida for discussion.
The work of M.H. was supported in part
by JSPS Research Fellowships for Young Scientists (\#15-10363).
That of K.T. was financially supported by 
Grant-in-Aid for Scientific Research (\#15740242).

\end{document}